# Light-induced quantum friction of carbon nanotubes in water

*Tanuja Kistwal[1+], Krishan Kanhaiya[3+], Adrian Buchmann[1], Chen Ma[1], Jana Nikolić[1], Julia Ackermann[2], Phillip Galonska[1], Sanjana S. Nalige[1], Martina Havenith[1*], Marialore Sulpizi[3*], Sebastian Kruss[1,2]\**

[1] Department of Chemistry and Biochemistry, Ruhr-University Bochum, Universitätsstraße 150, 44801 Bochum, Germany

[2] Fraunhofer Institute of Microelectronic Circuits and Systems, 47057 Duisburg, Germany.

[3] Department of Physics and Astronomy, Ruhr-University Bochum, Universitätsstraße 150, 44801 Bochum, Germany

+ These authors contributed equally

\* Correspondence: sebastian.kruss@rub.de; marialore.sulpizi@rub.de; martina.havenith@rub.de

## Abstract

Quantum friction describes the transfer of energy and momentum from electronically excited states in a material to a surrounding solvent. Here, we show that near-infrared (NIR) fluorescent single-walled carbon nanotubes (SWCNTs) exhibit quantum friction in water. The diffusion constants of functionalized SWCNTs in aqueous solution decrease linearly by around 50 % with increasing excitation power. In contrast, SWCNTs with quantum defects that localize excitons show no power-dependent diffusion. Chemical manipulation of exciton concentration by molecules that increase or decrease SWCNT fluorescence also modulate the diffusion constant by a factor of up to 2. Additionally, excitons increase the macroscopic viscosity of SWCNT solutions. Optical pump Terahertz (THz) probe spectroscopy reveals transient absorption features of water (37 cm⁻¹ and above 80 cm⁻¹), indicating energy dissipation into translational modes of its hydrogen bond network. Molecular dynamics simulations further support a mechanism in which exciton-induced dipoles enhance frictional forces. These findings establish that excitons in SWCNTs induce quantum friction in water.

## Introduction

Friction is a well-known phenomenon, with the first quantitative description dating back to Leonardo da Vinci (Amonton's law). To move an object over a surface, a force proportional to the normal force (weight) is necessary. On the nanoscale, friction becomes more complex due to surface topography.[1] A fundamentally different type of friction has been theoretically proposed and coined quantum friction.[2] It includes the non-adiabatic coupling between collective modes of solvent (water) dipoles and electronic modes specific to materials such as graphene or carbon nanotubes (CNTs).[2–4]

Subsequent experiments demonstrated that the resonance between graphene surface plasmon modes and water charge fluctuations (libration modes in the Terahertz (THz) region) cool hot electrons in graphene.[5] The water–graphene quantum friction force is negligible if the graphene electrons are at rest but becomes significant if they are driven at high velocity by phonons or an applied voltage.[6] In this context, THz spectroscopy serves as a powerful tool for probing distinct water populations within hydration shells,[7] effectively capturing dynamic solute-solvent interactions, including those driven by charge fluctuations.[4] Notably, these charge fluctuation-driven interactions extend beyond the primary hydration shell,[8] influencing broader solvation dynamics. Highly sensitive THz absorption studies have revealed the presence of two distinct water populations within the hydration shell, attributed to hydrophobic and hydrophilic hydration.[9] Simulations offer further insights into the anomalously high water friction on graphite, which is attributed to its THz plasmon mode, as well as the unique slippage behavior observed in CNTs.[2,3] Additionally, the water-carbon interface demonstrates remarkable thermal boundary conductance, exceeding that of other solid-liquid systems. This emphasizes the vital role of plasmon-hydron coupling and THz frequency modes for properties of graphene-based materials.[5]

Semiconducting single-walled carbon nanotubes (SWCNTs) are one-dimensional nanomaterials that fluoresce in the near-infrared (NIR) tissue transparency window.[10,11] Their fluorescence is best described by electron/hole pairs called excitons,[12] which diffuse along the axis of the SWCNTs for around 100 nm.[13,14] Because the size of excitons is larger than the diameter of SWCNTs, they are significantly affected by changes in the surrounding dielectric environment caused by bundling,[15] surfactants,[16] or DNA wrapping.[17] Unlike typical organic dyes, SWCNTs do not bleach and they do not blink like quantum dots (QDs).[11,18] SWCNTs are inherently hydrophobic,[19] but adsorption of surfactants, peptides proteins,[20,21] or π-stacking of nucleic acids [22] makes them water soluble. Another way to change their surface chemistry is covalent functionalization, which introduces a low number of σ-bonds into the sp² hybridized carbon lattice (sp³ quantum defects). They act as local traps for excitons and create novel photophysics.[23–26]

The optoelectronic properties of SWCNTs are highly sensitive to their chemical environment, which makes them ideal building blocks for (bio)sensors.[19] SWCNTs can be chemically tailored to change specifically their fluorescence in response to analytes. The mechanism has been attributed to conformational changes and changes in local solvation.[4] Such sensors have been used to image chemical signaling by cells,[27,28] for cancer or pathogen diagnostics,[29,30] or to image plant stress.[31,32]

To study fluorescent nanomaterials single particle methods such as fluorescence correlation spectroscopy (FCS) can be used.[33] FCS exploits the fluctuations in photoluminescence (PL) intensity emitted by freely diffusing particles within a confocal excitation volume of ~1 femtoliter (fl) in solution. It provides a temporal resolution of < 0.1 µs and yields insights into diffusion and photophysical phenomena.[34] Additionally, fluorescence fluctuations of nanomaterials are influenced by factors such as the corona layer or the presence of analytes.[35]

Here, we measure the diffusion of SWCNTs to study if excitons affect quantum friction. We use physical manipulation by changing (light) excitation, and chemical control by adding analytes or changing surface chemistry to identify the role of excitons in friction. We complement these experiments by studying the energy dissipation pathway via THz spectroscopy and molecular dynamics simulations.

## Results and discussion

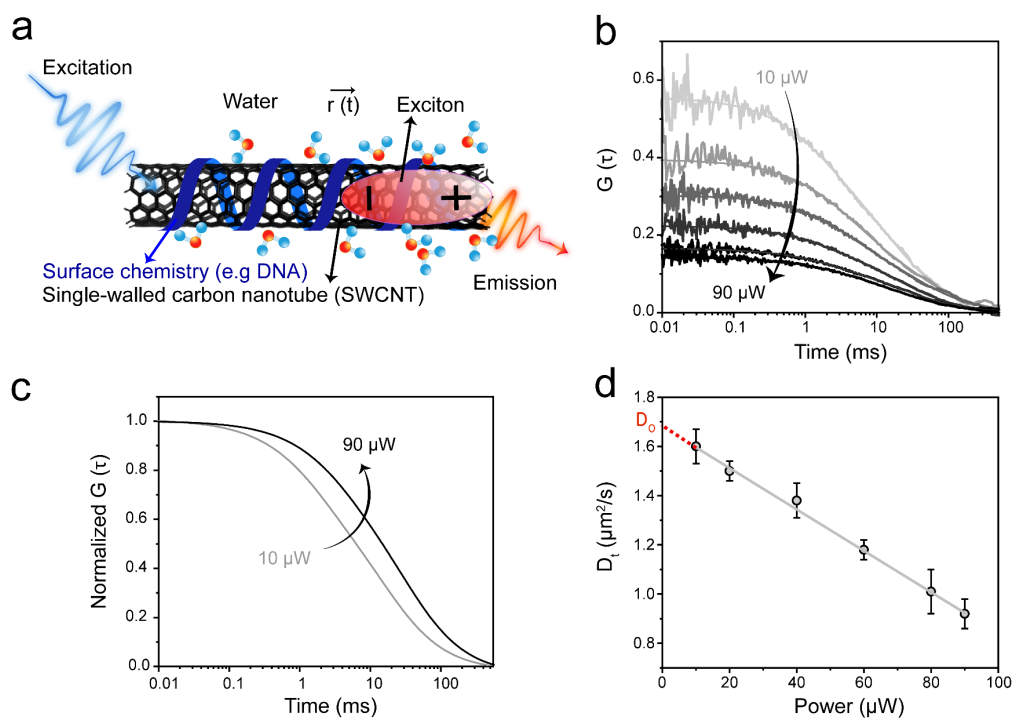

*Figure 1. Light-induced friction of carbon nanotubes in water. a) Schematic of experimental design. Single-walled carbon nanotubes (SWCNTs) modified with a biopolymer such as DNA are water-soluble. They are optically excited, which creates excitons that decay by either emitting near-infrared (NIR) photons or dissipating energy. The SWCNTs diffuse due to Brownian motion. b) Fluorescence correlation spectroscopy (FCS) of $(GT)_{10}$-SWCNTs shows an excitation power-dependent change of the fluorescence (> 900 nm) autocorrelation function ($\lambda_{exc}$ = 480 nm). c) The normalized and fitted (see equation 4, Table ST1) autocorrelation functions indicate slower diffusion with increasing power. d) Corresponding diffusion constants of $(GT)_{10}$-SWCNTs at different excitation power (n=3, mean ± SD). $D_0$ represents the translational diffusion constant ($D_t$) extrapolated (red dashed line) to zero excitation power from the linear fit (grey) of the data ($R^2$ = 0.996).*

We conducted single molecule fluorescence measurements to explore the diffusion behavior of SWCNTs in water under light excitation (Figure 1a). For this purpose, the hydrophobic SWCNTs (mainly semiconducting (6,5)-chirality) were functionalized with single stranded DNA ($(GT)_{10}$) or surfactants (deoxycholate (DOC), sodium cholate (SC), sodium dodecyl benzene sulphonate (SDBS)). We also prepared SWCNTs with additional Nitro-Aryl $sp^3$ quantum defects, which trap and localize excitons.[24,26] These samples were colloidally stable in aqueous solution as observed in absorbance (Figure S1a), one-dimensional (1D) fluorescence (Figure S1b,c), and two-dimensional (2D) (Figure S2) excitation-emission spectra.

In NIR FCS measurements we observed a decrease in the initial fluorescence autocorrelation amplitude $G(0)$ for increasing laser power (ranging from 10-90 µW) for $(GT)_{10}$-SWCNTs (Figure 1b), which can be expected for a NIR fluorophore of the length of a SWCNT (600 nm) with low quantum yield.[19] The normalized and fitted (Equation 4) correlation amplitude at low and high laser power (Figure 1c) showed that diffusion slowed down with higher power. By fitting the diffusion constants at different laser powers with a linear equation, we extrapolated the translational diffusion constant ($D_t$) value (1.7 µm²/s) for $(GT)_{10}$-SWCNTs at zero power (Figure 1d). This value is comparable to previously reported diffusion constants from 0.4 to 2.3 µm²/s, depending on the conditions and the lengths of the SWCNTs.[36,37]

A control experiment under identical conditions was conducted with the dye Atto 488 (Figure S3a) and showed a slight decrease in $G(0)$ value but no change of the normalized autocorrelation functions and the diffusion time (Figure S3b) under the same experimental conditions. It rules out effects from sample heating, which is known for surface-immobilized emitters.[38,39] Brightness of $(GT)_{10}$-SWCNTs increased also linearly with laser power (Figure S4a), indicating the absence of non-linear effects such as exciton-exciton annihilation. The increase in the number of (apparent) fluorescent particles (Figure 1b) from 3 to 9.3 in the confocal volume with higher laser power (Table ST1) can be attributed to the relatively small quantum yield of NIR fluorophores such as SWCNTs,[40,41] which means that they are not saturated by excitation. In addition, the diffusion behavior of $(GT)_{10}$-SWCNTs could be reversibly switched by changing the excitation power (Figure S4b). These results suggest that the enhanced friction by light excitation results from coupling between excitons and water molecules in the solvation layer of SWCNTs. This coupling leads to small energy and momentum losses (i.e. quantum friction). Consequently, a drag force slows down the Brownian motion of SWCNTs.

Next, we changed the solvent from water ($H_2O$) to heavy water ($D_2O$) (only solvents that provide colloidal stability are possible). $(GT)_{10}$-SWCNTs in a $D_2O$ based PBS buffer did show only a small power-dependent diffusion behavior (Figure S5a, Table ST2). This is in line with more pronounced cooling dynamics between graphene plasmon and water libration modes compared to other solvents such as $D_2O$ or methanol.[5] We also investigated a 80 % glycerol/water mixture and observed almost no changes in diffusion (Figure S5b, Table ST3). In contrast, for a 20% glycerol/water mixture we observed power-dependent changes in the diffusion constant similar to those of water. (Figure S5c, Table ST4). When glycerol content exceeds 40%, most water molecules will bond to glycerol through hydrogen bonding, and thus the THz spectrum of the solvent changes, which highlights that coupling to THz modes plays an important role.[42]

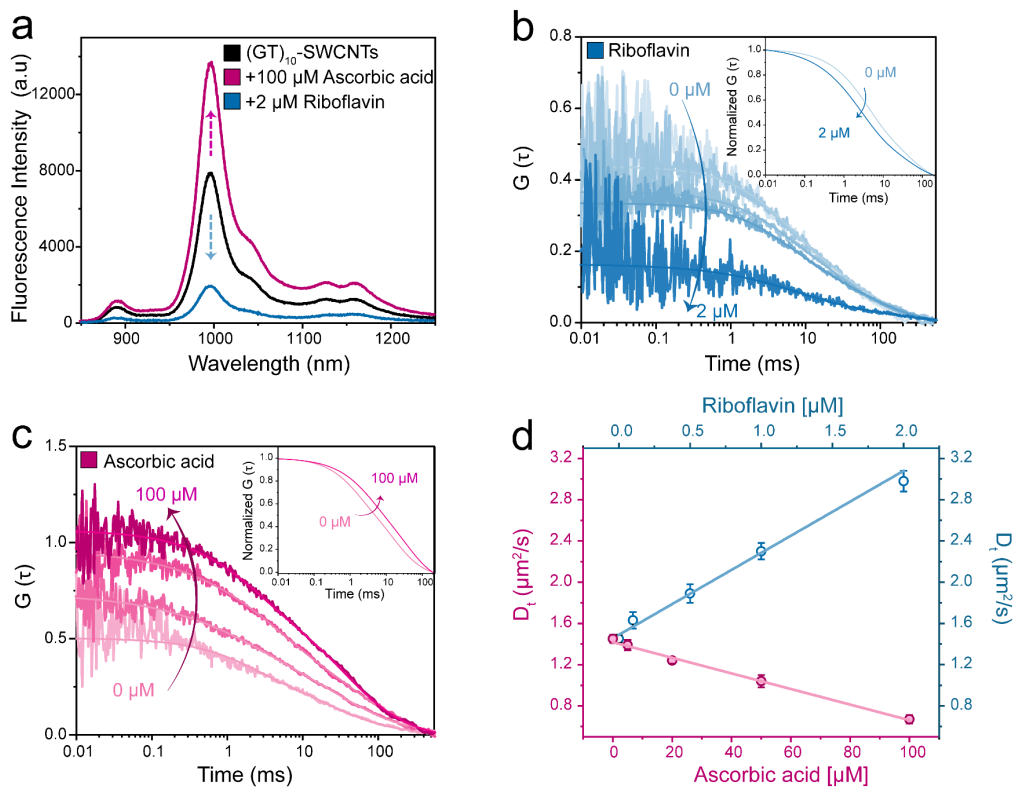

*Figure 2. Chemical manipulation of quantum friction.* a) Fluorescence spectrum of $(GT)_{10}$-SWCNTs without an analyte (black), with 100 µM ascorbic acid (deep pink) and 2 µM riboflavin (sapphire blue) in PBS buffer. b) Riboflavin (0 µM – 2 µM) and c) ascorbic acid (0 µM – 100 µM) change the autocorrelation curves of $(GT)_{10}$-SWCNTs in a concentration-dependent manner. The insets show the normalized and fitted autocorrelation function. d) Diffusion constants of $(GT)_{10}$-SWCNTs in response to different concentrations of ascorbic acid and riboflavin (n=3, mean ± SD). Fits = linear equations ($R^2$ = 0.98 for ascorbic acid and $R^2$ = 0.99 for riboflavin).

Increasing the light excitation is one way to change the exciton concentration. Another way is to change how fast or efficiently excitons decay. This mechanism is the basis for molecular (bio)sensors based on SWCNTs.[19] Thus, we added analytes that change the fluorescence quantum yield to study the chemical manipulation of quantum friction. The fluorescence of $(GT)_{10}$-SWCNTs increases in the presence of ascorbic acid and decreases in the presence of riboflavin (Figure 2a). These changes in fluorescence are anti-correlated to changes in THz absorption, which suggests a coupling of charge fluctuations in SWCNTs to charge density fluctuations in the hydration layer.[4] In FCS measurements riboflavin decreased the autocorrelation amplitude (Figure 2b), shifted the normalized autocorrelation curves (Figure 2b, inset), and reduced the diffusion time (Table ST5). On the other hand, ascorbic acid slowed down the diffusion of SWCNTs and increased the diffusion time (Figure 2c, Table ST6). For different analyte concentrations, diffusion increased linearly for analytes which decreased the quantum yield and *vice versa* (Figure 2d). In both cases, the diffusion constants changed by a factor of around 2 (Figure 2d, Table ST5 and ST6). Thus, chemical manipulation affects quantum friction the same way it affects exciton concentration (quantum yield). These results show that the change in diffusion is not an optical artefact and different to trapping of objects by light with optical tweezers.[43]

The next question was how the direct chemical environment/organic corona around a SWCNT affects quantum friction. In the case of DOC functionalized SWCNTs (DOC-SWCNTs) we observed a similar trend in power dependency as for (GT)$_{10}$-SWCNTs (Figure S6a), and the diffusion time increased (Table ST7). In contrast, DOC-SWCNTs with Nitro-Aryl sp$^3$ defects did not change their diffusion with excitation power (Figure 3a, Table ST8). It implies that a trapped/localized (non-moving) exciton does not affect friction. Surfactants like SC and SDBS change the chemical environment seen by the SWCNT. They exhibited the same power-dependent trend as when DNA was used for functionalization (Figure S6b,c). However, different coronas should affect as well the hydrodynamic radius as well, and indeed diffusion was faster for DNA and SDBS coated SWCNTs compared to DOC and SC (Figure S6d).

To disentangle the effects of hydrodynamic radius and quantum friction we determined the power-dependent diffusion constants for different types of functionalized SWCNTs (Figure S7a). The extrapolated diffusion constants at zero power represent the impact of the hydrodynamic radius and SWCNTs length on diffusion. In contrast, the normalized excitation power-dependent changes in diffusion reflect quantum friction (Figure S7b). This normalized change of the diffusion constants correlated with the SWCNTs emission wavelength (Figure 3b). A redshift in fluorescence emission is expected if the SWCNT is exposed to more water.[44] Therefore, a less dense corona (DOC, DNA) allows the exciton to 'see' more water, which causes more quantum friction. Another question that arises is how SWCNT chirality and thus curvature affects quantum friction or coupling. We extrapolated the diffusion constant and determined the value at zero excitation power ($D_0$) to be 1.00 µm$^2$/s for DOC-(6,4)-SWCNTs and 0.97 µm$^2$/s for DOC-SWCNTs (with mainly (6,5)-SWCNTs), which shows that for these SWCNT chiralities, there is no strong difference in quantum friction (Figure S8a, b).

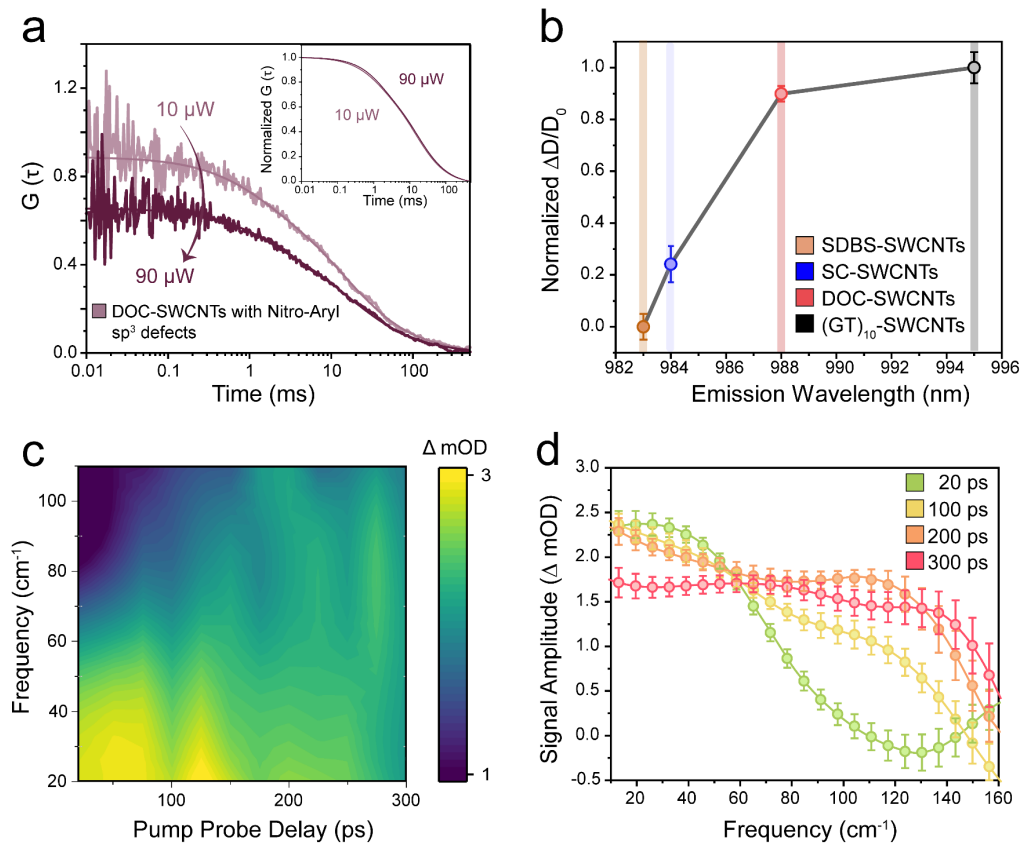

*Figure 3.* **The organic corona shields quantum friction and excitons change THz absorption by water.** *a) Excitation power-dependent fluorescence autocorrelation functions of DOC functionalized SWCNTs (DOC-SWCNTs) with Nitro-Aryl sp$^3$ quantum defects (wine) that serve as exciton traps show no power-dependent diffusion. The inset shows the normalized and fitted autocorrelation function. b) Normalized diffusion constant changes increase with emission wavelength for different functionalized SWCNTs (mean ± SD, n=3). Lines serve as visual guides. c) Optical Pump Terahertz Probe (OPTP) measurements of an aqueous DOC-SWCNTs solution. The difference in THz absorbance (Δ mOD) with excitation versus without exciton excitation (400 nm) is plotted against pump-probe delays. Positive values correspond to more absorption. d) Vertical slices at representative pump-probe time delays (mean ± SE, n=22).*

To understand the coupling between excitons and water, we then conducted Optical Pump Terahertz Probe (OPTP)[45] measurements of DOC-SWCNTs in water with 50 fs (400 nm) pump laser pulses at a fluence of 200 mJ/cm². On all spectral ranges and time scales we observed an increase in THz absorption, which was attributed to an increase in the polarizability of water (Figure 3c,d). After the optical excitation of the excitons, i.e. at a pump-probe delay time of Δt = 0.25 ps, we observed a spectral increase in absorption centered at approximately 50 cm$^{-1}$. For larger delay times the maximum absorption shifted towards lower frequencies centered around 37 cm$^{-1}$ (Figure S9a, b). Please note, that for further analysis we did not take into account signals at delay times of less than 0.3 ps, since these OPTP signals might suffer from artifacts due to pump-probe overlap. We fitted the signals with a Gaussian profile to extract the amplitude, which decays exponentially with a time constant of 0.71 ± 0.24 ps (Figure S9c). We also plotted the change in optical density (Δ mOD) for three pump fluences at 2 ps (Figure S9d). The results showed a linear increase in amplitude around 50 $cm^{-1}$ with higher pump power, in line with a coupling between the low-frequency modes of the exciton and those of the hydration water. In contrast, multiphoton processes would show a nonlinear increase with pump power.

At even longer delay times, we observed an increase in absorbance at higher frequencies above 80 $cm^{-1}$ (Figure 3c, d). It is known that water has collective translational modes (intermolecular hydrogen bond stretch) in this frequency range (centered ~80 $cm^{-1}$).[46] Thus, we propose that this indicates a coupling between excitons and the water's collective translational modes. The intermolecular stretching frequency in water has a maximum absorption of around 160-190 $cm^{-1}$, which is outside of our experimental accessible frequency range. Overall, our data suggests that excitons couple to translational water modes in the THz range, thereby causing momentum transfer.

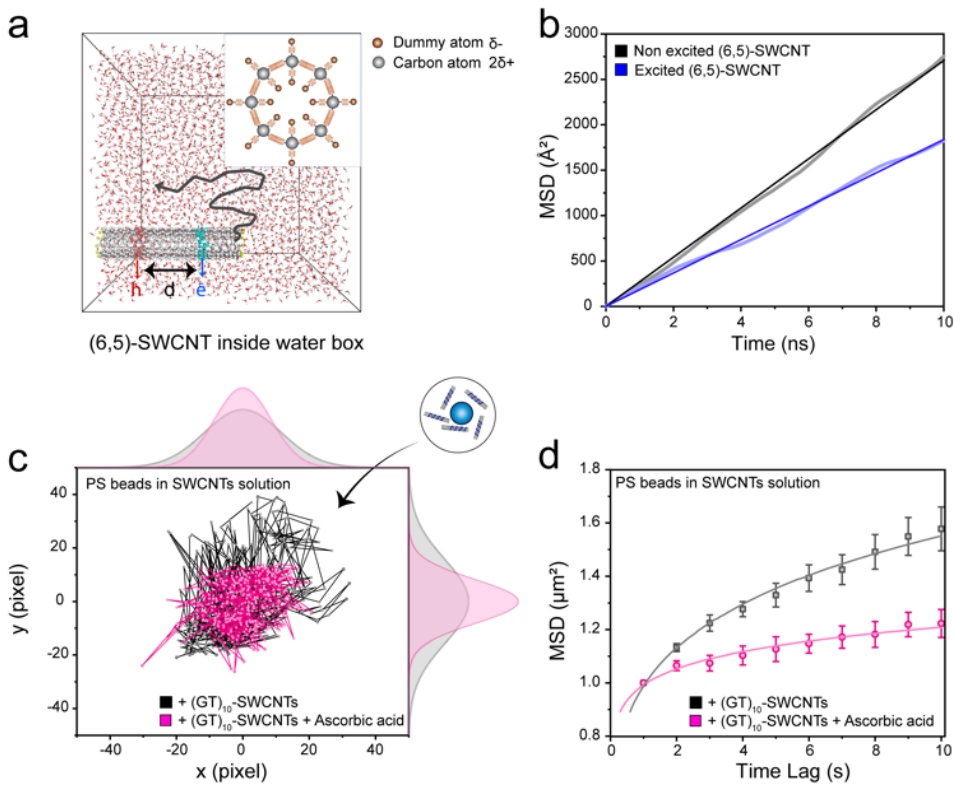

*Figure 4. Excitons change diffusivity and affect macroscopic viscosity.* a) Molecular dynamics (MD) simulation of excited SWCNTs. A (6,5)-SWCNT (length 4.1 nm) is allowed to move freely (black trace) in a three-dimensional periodic water box. The ends of the SWCNT are capped with hydrogens (yellow). Excitons are modelled by placing charges in the blue (e, electron) and red (h, hole) annular regions separated by the distance d. Inset: Schematic of the SWCNT cross-section. For each carbon atom (grey, charge $2\delta+$) two dummy/virtual atoms (brown, charge $\delta-$ each) are placed at a distance of 0.6 Å to mimic the electrostatics of a π-electron cloud. The virtual atoms provide polarizability in response to solvent and ion dynamics. b) Mean square displacement (MSD) curves (semi-transparent) and linear fits for both excited (with excitons) and non-excited (without excitons) SWCNTs (d=2 nm). c) Single particle trajectories of (fluorescent) polystyrene (PS) beads (5 µm) diffusing in a PBS buffer solution of $(GT)_{10}$-SWCNTs with and without ascorbic acid (100 µM), which increases exciton concentration (see Figure 2a). Gaussian histograms show the distribution of x and y positions. d) Ensemble time-averaged MSD plots for PS beads indicate slower diffusion in the presence of more excitons (+ascorbic acid) (mean ± SE, n=4 independent experiments with > 20 PS beads each).

We next used molecular dynamics (MD) simulations to investigate how a delocalized electron-hole pair (i.e. exciton) influences the interaction between finite-length SWCNTs and water. For this purpose, we simulated the diffusion of semiconducting (6,5)-SWCNTs (4.1 nm in length, capped with hydrogen atoms) in about 14000 water molecules at 298 K and 1 atm under isothermal and isobaric (NPT) conditions (Figure 4a). The (6,5)-SWCNTs are described by a new classical polarizable model[47,48] ((Figure S10 a-d and Table

ST14), which is capable of reproducing friction at the wall/water interface at the accuracy level of atomistic calculations including electronic structure (Figure S10e). The formation of an electron-hole pair (exciton) upon optical excitation is described by the addition of positive and negative delocalized charges with a varying spatial separation of 1-2 nm (expected size of the exciton).[14] These classical electrons and holes are obtained by adjusting the charges of dummy atoms attached to the carbon atoms (Drude-like oscillators that mimic the carbon π orbitals in our model) thus creating a charge in the blue annular region (electron) and decreasing it in the red annular region (hole) by 0.005e(Figure 4a). Each annular region contained 44 dummy atoms, resulting in dipole moments of 2.2 eÅ - 4.4 eÅ for separation distances (d) between electron and hole of d = 1 nm and d = 2 nm, which interact with the surrounding water.

The diffusion constant (D = slope/6), calculated from the slope of the linear fit to the mean square displacement (MSD) (Figure 4b), decreased by > 30 % from 450 µm²/s (non-excited) to 307 µm²/s (excited). Notably, in experiments excitons are mobile and the SWCNTs lengths used are much longer than those in our simulations. However, the numbers are in good agreement if we consider the length dependence of translational diffusion.[49] Therefore, the MD simulations validated the experimental findings.

Finally, we wanted to understand if quantum friction can affect the macroscopic viscosity of a liquid. Thus, we conducted single particle tracking experiments. Tracking experimentally single SWCNTs in wide-field microscopy proved challenging due to their rapid diffusion and we could not use glycerol to slow it down as discussed above (Figure S5b). As an alternative, we imaged Rhodamine B (RhB)-labeled polystyrene (PS) beads (5 µm diameter) in a solution with SWCNTs (with and without ascorbic acid to change quantum friction). The MSDs of the trajectories (Figure 4c, figure S11a) were fitted to a power law (MSD $(t) \propto t^\alpha$) and indicated subdiffusion ($\alpha < 1$). We found that the PS beads moved faster without ascorbic acid (Figure 4d, figure S11b). Ascorbic acid could also change the hydrodynamic radius or the viscosity of the solution but at such low concentrations, this is very unlikely. These results indicate that light-induced quantum friction increases the apparent macroscopic viscosity experienced by other objects in an aqueous solution of SWCNTs.

## Conclusion

Our study demonstrates that quantum friction between SWCNTs and water exists and is affected by light excitation, interface chemistry, and environmental factors. Exciton-water coupling generates a drag force that slows down the Brownian motion of SWCNTs. Since Einstein's seminal work on diffusion, we know that thermal energy is linked to the stochastic movements of Brownian motion. Light-induced quantum friction represents an additional factor and under our experimental conditions, it changes diffusion constants by a factor up to 2. We demonstrated that either physical (light intensity) or chemical (biomolecules that change fluorescence quantum yield) manipulation changes quantum friction (Figures 1 and 2). This observation pinpoints to a central role of the exciton concentration. Additionally, the mobility of the exciton (Figure 3a) and the organic corona between the carbon lattice and water (Figure 3b) determined the magnitude of quantum friction. THz OPTP measurements provided insights into the microscopic origins of this phenomenon and indicated energy/momentum transfer from the exciton to translational and stretching modes of hydrogen bonds in water.[4] MD simulations of SWCNT diffusion and exciton dipole coupling to the solvent further confirmed an increase in friction (Figure 4a,b). Excited SWCNTs even impact the macroscopic viscosity experienced by micrometer-sized beads in water (Figure 4c, d).

These findings show that quantum friction directly affects the motion of a nanoscale object and that there is momentum transfer. It is different from laser traps or optical tweezers because one can chemically change it (Figure 2d) and it takes place at much lower laser intensities.[50] Our study experimentally demonstrates that quantum friction i.e. light-induced slowing down of Brownian motion exists. It can be physically or chemically manipulated to affect the Brownian movement of a nanomaterial in an aqueous solution or change the apparent viscosity of a solution. Excitons interact with the solvent, which slows down the movement of the object. Given the anisotropy of SWCNTs, one can anticipate approaches to control the movement of microswimmers or nanorobots. Additionally, it raises the question if other materials that have similar excited state dynamics also exhibit quantum friction in water or other solvents.

## Materials and Methods

### Single-walled carbon nanotube preparation

If not stated otherwise all chemicals were purchased from Sigma Aldrich (Germany). Unless specifically stated, all experiments were performed with (6,5)-enriched SWCNTs (Sigma Aldrich, Signis® SG65i, CoMoCAT™ synthesis technology). For DNA functionalized SWCNTs 150 μL of 2 mg/mL single stranded DNA (e.g. $(GT)_{10}$ ) in 1x phosphate-buffered saline buffer (PBS, pH 7.4) was mixed with 75 μL of 2 mg/mL SWCNT in PBS and 75 μL PBS, followed by tip sonication (Fisher Scientific, FB120, 120W, amplitude 35%, 9 s pulse on and 1 s off, 15 min). The obtained solution was centrifuged for 30 min at maximum speed (21k g), the supernatant was collected, and this procedure was repeated two more times. The final supernatant was stored at 4°C until further experiments were performed. For $(GT)_{10}$-SWCNTs experiments in $D_2O$, the PBS buffer was prepared with $D_2O$ instead of $H_2O$.

The separation of (6,4)-SWCNTs was performed according to the following protocol.[51] (DOC)-SWCNTs were mixed with Polyethylene glycol (PEG) (MW 6 kDa, 8% w/v), dextran (Carl Roth, MW 70 kDa, 4% w/v), and the surfactants DOC (0.025% w/v), SDS (0.5% w/w), and SC (ranging from 0.5% to 0.9% w/w in 0.1% increments). The SWCNTs chiralities in the two phases could be adjusted by adding HCl. Then, a one-step approach was used by adding a specific volume of HCl (Hydrogen chloride) and NaClO (Sodium hypochlorite) with 10–15% available chlorine for pH-driven and electronic separation, allowing the collection of monochiral (6,4)-SWCNTs in the bottom phase (B3). The solution was then dialyzed (using a 300 kDa dialysis bag, Spectra/Por, Spectrum Laboratories Inc.) against a 1% DOC solution to remove dextran and obtain a stable 1% DOC-(6,4)-SWCNT solution.

For DOC-SWCNTs, 150 µL 2% (m/v) DOC in $H_2O$ was mixed with 150 µL of 2 mg/mL SWCNTs (in $H_2O$), followed by tip sonication and centrifugation similar to the conditions for $(GT)_{10}$-SWCNTs preparation. The acquired supernatant was stored at 4℃. Sodium dodecyl benzene sulphonate (SDBS) and SC functionalized SWCNTs were prepared according to the same procedure as DOC-SWCNTs.

Quantum defect introduction was performed according to a previously developed protocol.[52] Briefly, 20 µL of 4 mM 4-nitrobenzol diazonium tetrafluoroborate diazonium salt (dissolved in water) was added to 20 mL of 10 nM SDBS-SWCNTs solution. Then the mixture was irradiated with green light (550 nm) while stirring for 15 min. The obtained solution was mixed with the same volume of acetonitrile (ACN), and consequently, the SWCNTs precipitated. The pellet was then washed with $H_2O$ 2-3 times to remove residual SDBS and ACN. Finally, the acquired precipitate was redispersed in 1% DOC by 15 min tip sonication followed by centrifugation for 30 min at 21000g. The collected supernatant was used for the experiments. The length of SWCNTs prepared by this procedure is ~ 600 nm.[26]

**NIR spectroscopy**

**1D fluorescence spectra**

1D spectra of 0.5 nM $(GT)_{10}$-SWCNTs with or without analytes (2 µM riboflavin and 100 µM ascorbic acid in aqueous solution) or 0.5 nM DOC, SC, SDBS functionalized SWCNTs were measured in a custom-built setup based on an Olympus IX73 microscope and a solid-state laser (Quantum gem-561, 561 nm). The emission spectra were captured with an Andor iDus InGaAs 491 array NIR detector coupled to a Shamrock 193i spectrometer (Andor Technology Ltd., Belfast, Northern Ireland).

**2D fluorescence spectra**

The same setup as for 1D spectra was used. However, to obtain 2D excitation-emission spectra of 2 nM SWCNTs in various surfactants and $(GT)_{10}$-SWCNTs in 1xPBS ($D_2O$) at pH 7.4, a monochromator (MSH150) equipped with a LSE341 light source (LOT-Quantum Design GmbH, Germany) was employed for tunable excitation.

**FCS measurements**

FCS measurements were performed with a MicroTime 200 system (PicoQuant, Germany), equipped with pulsed lasers at 485 nm (LDH-C-D-485) and 530 nm (LDH-D-TA-530), an Olympus IX73 inverted confocal laser scanning microscope equipped with a 60x water objective (Olympus, NA 1.2, UPlanSApo), and single-photon avalanche photodiodes (SPADs) detectors (Excelitas Technologies, Canada). Samples at a concentration of 1 nM were excited with a pulsed laser at 485 nm, operating at a frequency of 40 MHz. DOC-(6,4)-SWCNTs showed weak emission when excited at 480 nm. Consequently, we used for it the 532 nm excitation. Since the 532 nm laser could not achieve higher power levels in pulsed mode, we used continuous wave excitation at 532 nm for this measurement, ensuring that the excitation power remained consistent. The emitted light was separated from the excitation light through a dichroic mirror (R405/488/532/635, Semrock, USA), passed through a 900 nm long pass filter (Thorlabs) to block the excitation light, and then focused onto a 50 µm pinhole and directed to the SPAD detectors. For DOC-(6,4)-SWCNTs, a 800 nm long pass filter (Thorlabs) was used. The refractive index and also viscosity corrections were done by adjusting the collar settings.[53]

The autocorrelation function of the fluorescence intensity I is defined as:

$$G(\tau) = \frac{<I(t)I(t+\tau)>}{<I(t)^2>} \qquad (1)$$

$G(\tau)$ correlates the fluctuation of a fluorophore's intensity at time $t$ and after time lag $\tau$. Fluctuations arise due to the diffusive motion of the fluorophore through the 3D Gaussian confocal volume having width $w_z, w_{xy}$. The correlation function corresponding to the diffusion is:

$$G_D(\tau) = \frac{1}{N}\left[1+\frac{\tau}{\tau_D}\right]^{-1}\left[1+\frac{\tau}{w^2\tau_D}\right]^{-\frac{1}{2}} \qquad (2)$$

$N$ is the total number of molecules in the confocal volume and $\tau_D$ is the diffusion time of that system. It is linked to the diffusion constant D by:

$$\tau_D = \frac{w_{xy}^2}{4D} \qquad (3)$$

The structural parameter $w = \frac{w_x}{w_{xy}}$ was calibrated using the known Atto 488 dye (1 nM) in water ($D_t$ = 400 µm$^2$/s).[54] The calculated excitation volume was 1.5 fl.

To analyze the FCS data the software Igor Pro 6.34A and the following equation was used for fitting:

$$G_D(\tau) = \frac{1}{N}\left[1+\frac{\tau}{\tau_D}\right]^{-1}\left[1+\frac{\tau}{w^2\tau_D}\right]^{-\frac{1}{2}}\left[1+\frac{T}{1-T}\exp\left(-\frac{\tau}{\tau_t}\right)^\beta\right] \qquad (4)$$

$T$ is the fraction of the fluorescent molecules in the dark state and $\tau_t$ signifies the corresponding lifetime. The stretching exponent $\beta$ is a marker for the degree of heterogeneity in the associated dynamics.[33]

**THz measurements**

The OPTP spectrometer was described in detail previously.[45] In summary, the system uses 50 fs, 800 nm laser pulses generated by a Ti: Sapphire amplified laser to produce a broadband THz probe pulse via a two-color air plasma filament.[55] Part of the 800 nm wavelength laser radiation is frequency-doubled in a (Beta Barium Borate) BBO crystal to generate 400 nm light, which serves as the optical pump. We measure the changes in THz absorption upon optical excitation as a function of pump-probe delay, Δt, which can be adjusted between 0.25 ps and 300 ps using a mechanical stage. To eliminate interference effects and the excitation of free charge carriers, we employed a window-less, free-flowing jet with a thickness of 20 µm as the sample.[45,56] A 80 ml solution of SWCNTs (~100 nM) was circulated in the jet for 96 hours. A defoaming agent (BYK 025) was added to the reservoir to prevent foam generation. This defoaming agent remains as a thin film on the surface and does not interfere with the sample measurements. The THz field is detected using electro-optic sampling (EOS) with a 100 µm thin Gallium phosphide (Ga P) crystal. For further analysis, the electric fields are Fourier transformed. The difference in THz transmission before and after optical excitation is expressed as Δ mOD. Positive values indicate a decrease in transmission upon optical excitation. The fluency of the blue light was varied from 50 mJ/cm² to 120 mJ/cm² and then to 200 mJ/cm². As a reference, we also measured pure water at a fluence of 200 mJ/cm². In the plots, we show data for a fluence of 200 mJ/cm² unless stated otherwise.

**Tracking of rhodamine B labeled polystyrene beads in a SWCNT solution**

5 µm PS beads (Sigma Aldrich product number: 79633) were used for particle tracking. First, 100 µL of the solution containing 1 part PS beads to 9 parts PBS 200 was pipetted out. Next, 30 µL of RhB dye (Sigma-Aldrich, CAS No: 81-88-9) was pipetted from a freshly prepared stock solution (1 mM) and added to the PS beads solution. Following this, 770 µL of PBS was added, bringing the final volume of the sample to 1 mL. After vertexing the mixture, the sample was incubated in the dark for 1 hour to allow the dye to adsorb onto the beads. The next step involved washing out the excess dye from the suspension by centrifuging the sample twice for 5 minutes each time.

For particle tracking microscopy, we used the Leica THUNDER Imager 3D Cell Culture & Infinity Laser Scanner, employing fluorescence as the contrast method with an excitation wavelength of 575 nm and filter cube CYR7101. A 63x oil immersion objective was utilized for imaging. The sample was placed in a glass petri dish, which was then covered with a glass slide. The total sample volume was 200 µL, consisting of 186 µL of RhB PS bead solution and 14 µL of 70 nM $(GT)_{10}$-SWCNTs with or without 1 µL of 20 mM ascorbic acid. Images were captured at a frame rate of 10 frames per second. After collecting the videos, we processed the data using a Python script based on the trackpy package. From the trajectories, we then calculated the ensemble time-averaged mean squared displacement (MSD).[57]

**Computation of Friction and Diffusion in Water**

The classical atomistic MD simulations were run with a source Large-scale Atomic/Molecular Massively Parallel Simulator (LAMMPS)[58] to estimate the interfacial friction coefficient and diffusion of graphene and (6,5)-SWCNTs in explicit water. The models for graphene slab (2.5x2.6 nm$^2$) with 1600 water molecules and (6,5)-SWCNTs (3x3x4.1 nm$^3$) with 1100 water molecules systems were created in Material Studio.[59] For the calculation of the interfacial friction the employed graphene system is periodic in the x-y directions and non-periodic in the direction perpendicular to the surface whereas all the carbon nanotube systems are 3D periodic with an infinite nanotube along the axial direction. Both nonpolar and polar systems are analyzed with harmonic Consistent Valence Forcefield (CVFF)[60] and Interface Force Field-CVFF (IFF-CVFF)[61] parameters respectively (Table ST14) which use 12-6 Lennard Jones (LJ) potential for the vdw interactions. In the non-polar model, the carbon atom C is neutral and only has LJ interactions with the water, whereas, in the polarizable model, each carbon is decorated with two flexible negatively charged dummy atoms which mimic the π orbitals and are perpendicular to the plane of C-atoms. The dummy atoms are connected by harmonic bonds and angles restraints (Table ST14 for parameters). A similar simple model to include the metal polarization, which consists of a Lennard-Jones potential and a harmonically coupled core–shell charge pair for every atom has been recently developed and has proved to reproduce the classical image potential of adsorbed ions as well as surface, bulk, and aqueous interfacial properties in agreement with experiments.[48] Here, two layers of virtual atoms sandwich the carbon layer in between to form a single graphene sheet or SWCNTs (Figure S10 a-d). The dummy atoms mimic the π electron cloud and add polarizability to the carbon atoms.[47] The polarizable carbon carries a partial positive charge (+2δ) and the two dummy atoms carry a negative half charge (-δ) so the overall C-atom is neutral. However, there is an additional dipole contribution to each carbon atom. Hence, polarizable graphene/SWCNTs also have a columbic interaction with the surrounding water.

The Green Kubo friction coefficient has been calculated to estimate the strength of the interfacial interaction of water with the graphitic surfaces (Figure S10e),[62,63] according to the formula:

$$\lambda_{GK} = \frac{1}{AnK_BT} \int_0^\infty <FL(t)FL(0)> \quad (5)$$

Where $A$ is the area of the surface, $n$ is the number of dimensions ($n$ = 2 for graphene and 1 for CNT), $K_B$ is the Boltzmann constant, $T$ is the temperature, $FL$ is the lateral force acting on the surface for graphene or the force along the axial direction for the CNT. The integral of the autocorrelation of $FL$ is used to compute the Green Kubo friction coefficient as per equation 5. The friction coefficient for nonpolar graphene and (6,5)-SWCNT was computed with the CVFF parameters and polar graphene and (6,5)-SWCNTs is computed with the IFF-CVFF polarizable model.[48,61]

We observe a higher friction coefficient of around 6.5x10$^4$ Ns/m$^3$ at the graphene interface with the polarizable model as compared to 2x10$^4$ Ns/m$^3$ for nonpolar graphene which is also the typical value observed with other force fields.[3] Notably, the value for the friction coefficient obtained with our polarizable model is in very good agreement with the *ab initio* estimates of 4.5x10$^4$ Ns/m$^3$ [63] and 9.5x10$^4$ Ns/m$^3$ [64] obtained with revPBE-D3 (Figure S10e) and optB88-vdw (Figure S10e) functional respectively. We also observed that the friction coefficient increases for water in contact with the external surface of SWCNTs from 6.5x10$^4$ Ns/m$^3$ for the non-polar model, to 15x10$^4$ Ns/m$^3$ for the polarizable. The result for the polarizable model is in good agreement with ab initio molecular dynamics results from a previous study.[64] Hence, the new polarizable model permits to reproduce electronic structure level accuracy at the cost of simple classical force filed simulations introducing the interaction of the polarizable electron cloud with the polar solvent. With the new and improved IFF-CVFF polarizable model, we also estimated the diffusion behavior of (6,5)-SWCNTs. Latest IFF-CVFF polarizable graphite model (Table ST14) have been validated with rigour by reproducing bulk properties such as density and bulk modulus and interfacial properties such as surface energy, hydration energy and water contact angle which are in excellent agreement with experimental observations and are suitable to model graphitic materials in various applications.

The diffusion constant was computed with a (6,5)-SWCNT of length 4.1 nm placed inside a cubical 3D periodic box of 140000 water molecules modeled with flexible SPC parameters (CVFF). After pre-equilibration of the simulation box in an isothermal and isobaric ensemble (NPT), the simulation trajectory was run for another 20 ns with a timestep of 0.5 fs with a coordinate snapshot being generated at every 1 ps. The (6,5)-SWCNT was end capped with hydrogen atoms and allowed to diffuse inside the box unconstrained with NPT ensemble at 298 K and 1 atm. Hydrogen parameters are borrowed from the CVFF models. The trajectory was analyzed to compute the mean square displacement (MSD) of the center of mass of the CNT with time. The slope (m) of the MSD vs time plot was used to evaluate the diffusion constant $D = m/6$. The diffusion constant was calculated for both non-excited and excited CNTs. The process of exciting the nanotube (in molecular dynamics with classical potential) by introducing an exciton via addition of an axial dipole along the nanotube is described in the main text.

## Acknowledgments

This work is funded by the Deutsche Forschungsgemeinschaft (DFG, German Research Foundation) under Germany's Excellence Strategy—EXC 2033–390677874—RESOLV. This work is further supported by the "Center for Solvation Science ZEMOS" funded by the German Federal Ministry of Education and Research BMBF and by the Ministry of Culture and Research of Nord Rhine-Westphalia. T.K. acknowledges a postdoc fellowship by the Humboldt Foundation. We thank J. Enderlein for fruitful discussions on FCS and A. Janshoff on particle tracking.

## Author contributions

Conceptualization and Experimental Design: S.K and T.K, Data Acquisition: T.K, C.M, J.N, and A.B, Funding Acquisition: S.K, M.S., M.H. Data Analysis and Interpretation: T.K, C.M, J.N, S.K, A.B, S.N, M.H, K.K, and M.S. Simulation: K.K and M.S with input from T.K. and S.K, A.B, M.H., THz measurement: A.B, S.N, M.H., Manuscript Writing: T.K, S.K, C.M, J.N, A.B, M.H, K.K, M.S. Review and editing: T.K, S.K, C.M, J.N, J.A, P.G, A.B, S.N, K.K, M.S and M.H.

## Competing interests

The authors declare no competing interests.

## Additional information

Supplementary information available.

# Supplementary information

# Light-induced quantum friction of carbon nanotubes in water

*Tanuja Kistwal[1+], Krishan Kanhaiya[3+], Adrian Buchmann[1], Chen Ma[1], Jana Nikolić[1], Julia Ackermann[2], Phillip Galonska[1], Sanjana S. Nalige[1], Martina Havenith[1*], Marialore Sulpizi[3*], Sebastian Kruss[1,2]\**

[1] Department of Chemistry and Biochemistry, Ruhr-University Bochum, Universitätsstraße 150, 44801 Bochum, Germany

[2] Fraunhofer Institute of Microelectronic Circuits and Systems, 47057 Duisburg, Germany.

[3] Department of Physics and Astronomy, Ruhr-University Bochum, Universitätsstraße 150, 44801 Bochum, Germany

+ These authors contributed equally

\* Correspondence: sebastian.kruss@rub.de; Marialore.Sulpizi@rub.de; martina.havenith@rub.de

**Table of Contents**

**Supplementary figures**

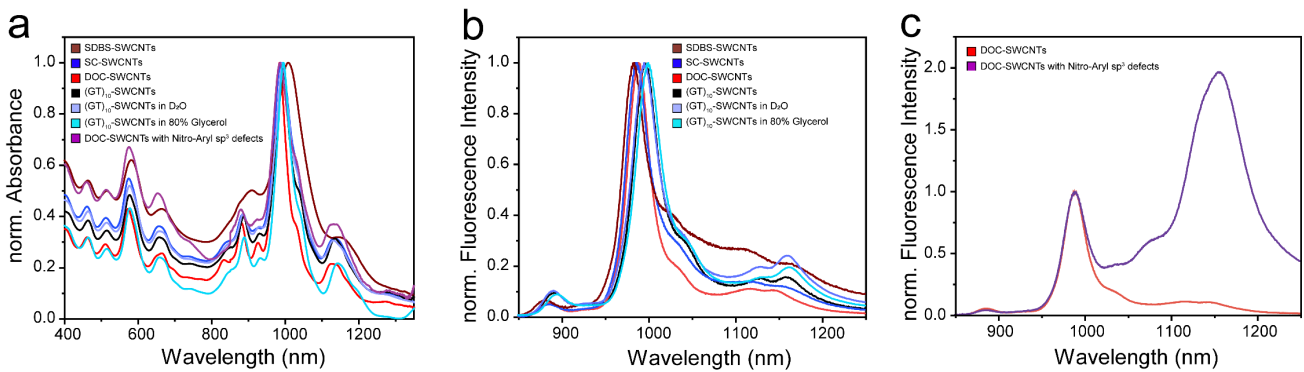

***Figure S1. Spectra of SWCNT samples in different surfactants***. *a) Normalized absorbance and b) normalized fluorescence spectra of SDBS functionalized SWCNTs (SDBS-SWCNTs) (brown), SC functionalized SWCNTs (SC-SWCNTs) (dark blue), DOC functionalized SWCNTs (DOC-SWCNTs) (red), DNA functionalized SWCNTs (GT)$_{10}$-SWCNTs in PBS buffer (black),(GT)$_{10}$-SWCNTs in D$_2$O PBS (neon blue), (GT)$_{10}$-SWCNTs in 80% glycerol (cyan). C) Emission spectra of DOC-SWCNTs with (wine) and without sp$^3$ quantum defects (red).*

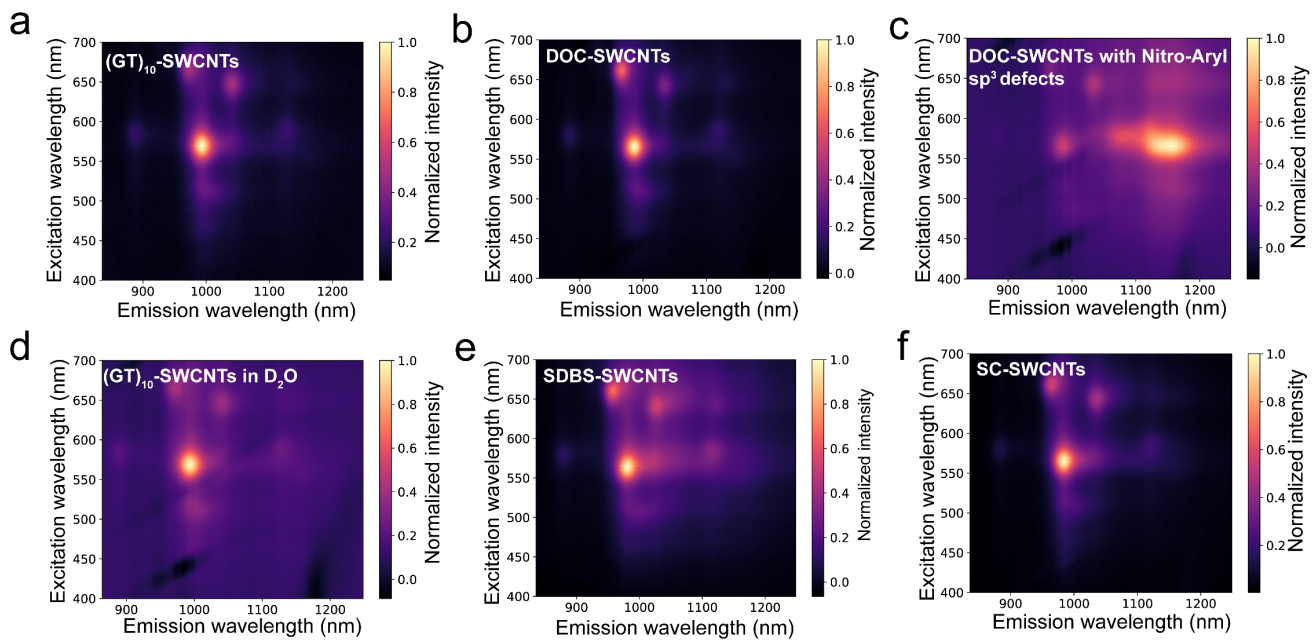

*Figure S2. 2D excitation-emission spectra of SWCNT samples.* *a) (GT)$_{10}$-SWCNTs, b) DOC-SWCNTs, c) DOC-SWCNTs with Nitro-Aryl sp$^3$ quantum defects, d) (GT)$_{10}$-SWCNTs in D$_2$O based PBS, e) SDBS-SWCNTs, f) SC-SWCNTs. The concentration of SWCNTs was 2 nM.*

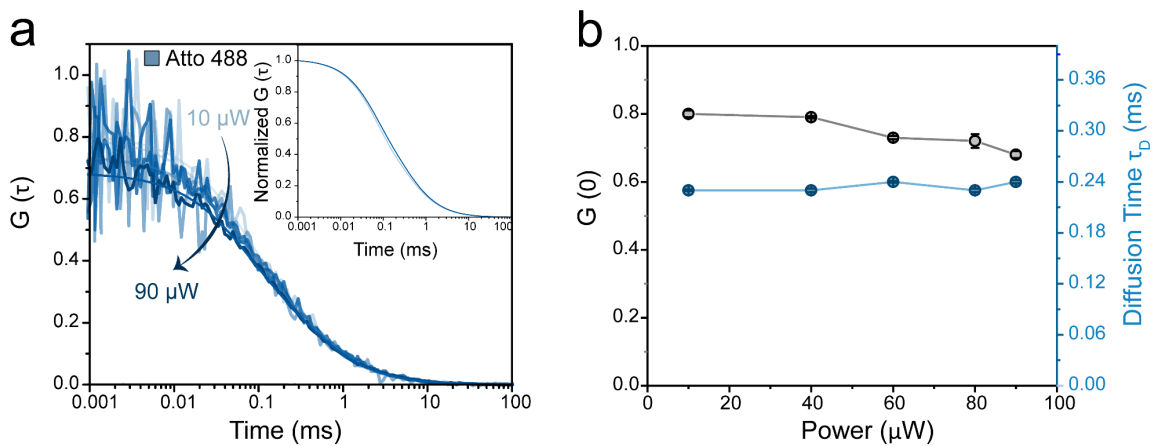

*Figure S3. FCS control experiment with Atto 488 dye.* *a) Fluorescence correlation curves of Atto 488 dye (1 nM) in water with increasing excitation power. b) Variation of initial correlation amplitude G (0) and diffusion time ($\tau_D$) as a function of excitation power (n=3, mean ± SD).*

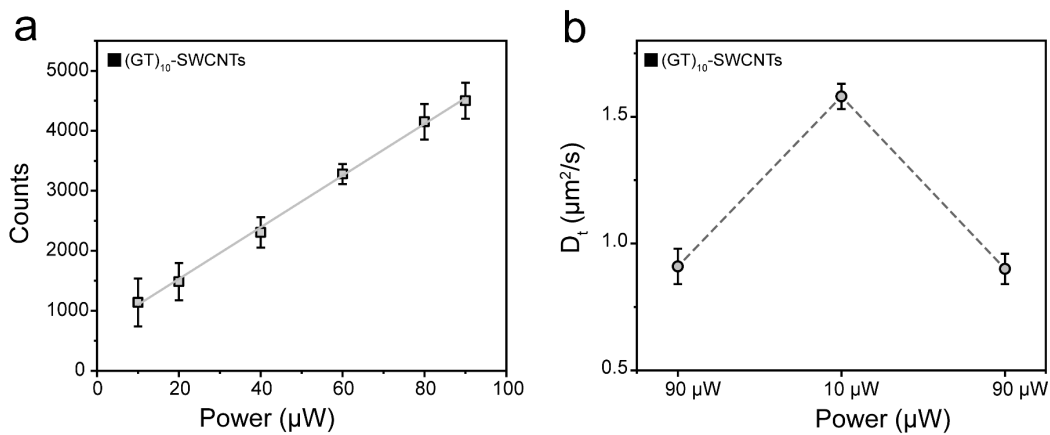

*Figure S4. Fluorescence intensity scales linearly with excitation power and the change of the diffusion constant is reversible.* Molecular brightness of $(GT)_{10}$-SWCNTs (1 nM) as a function of excitation power (Microtime 200 setup, $\lambda_{exc} = 480\ nm$), and fitted by a linear equation (grey) ($R^2 = 0.997$) (n=3, mean ± SD) b) The diffusion constant of $(GT)_{10}$-SWCNTs can be reversibly changed with excitation power (dashed grey lines show a visual representation) (n=3, mean ± SD).

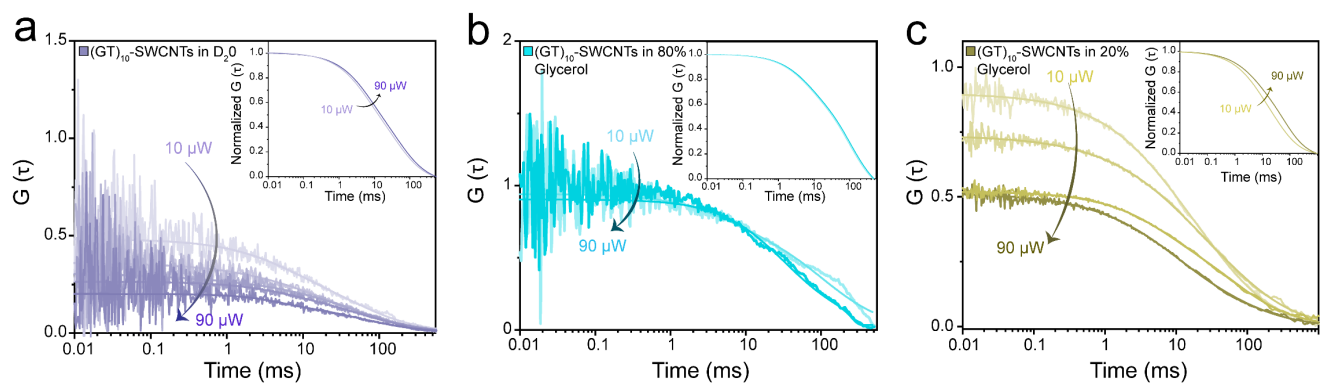

*Figure S5. Solvent effects on quantum friction.* Excitation power-dependent change (10 µW – 90 µW) of fluorescence autocorrelation functions of SWCNTs in different solvents. a) $(GT)_{10}$-SWCNTs in $D_2O$-based PBS (Table ST2), b) $(GT)_{10}$-SWCNTs in 80% glycerol in water (Table ST3), b) $(GT)_{10}$-SWCNTs in 20% glycerol in water (Table ST4).

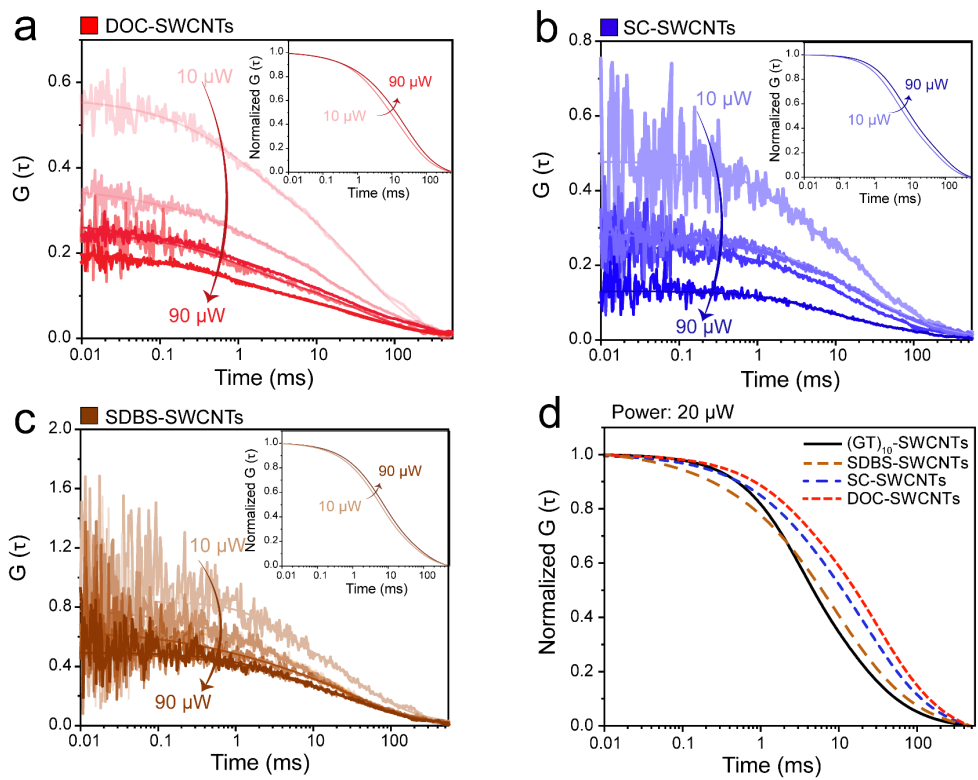

*Figure S6. The organic corona can shield quantum friction. Excitation power-dependent change (10 µW – 90 µW) of fluorescence autocorrelation functions of SWCNTs solubilized in different surfactants that create different organic coronas. The inset shows the normalized and fitted autocorrelation functions. See SI tables for fit functions. a) DOC-SWCNTs (red) b) SC-SWCNTs (blue) c) SDBS-SWCNTs (brown). d) Normalized autocorrelation functions of functionalized SWCNTs at 20 µW excitation power.*

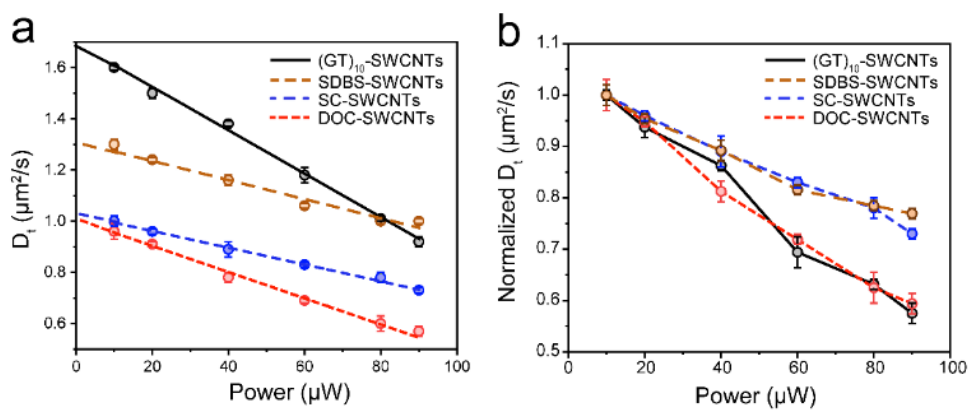

*Figure S7. Impact of surface chemistry/corona shielding on diffusion. a) Diffusion constants as a function of excitation power, the diffusion constants were fitted to a linear equation (Table ST12). b) Normalized diffusion constants as a function of power for $(GT)_{10}$-SWCNTs (black), SDBS-SWCNTs (brown), SC-SWCNTs (blue), and DOC-SWCNTs (red).*

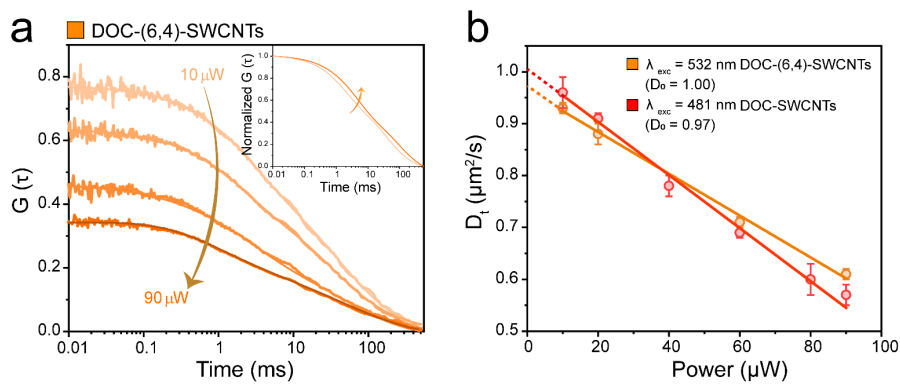

*Figure S8. Comparison between DOC-(6,4)-SWCNTs and DOC-(6,5)-SWCNTs. a) Fluorescence autocorrelation curves of DOC-(6,4)-SWCNTs (inset shows the normalized and fitted autocorrelation function) for an increasing excitation power. b) Diffusion constants of DOC-(6,4)-SWCNTs (orange) and DOC-SWCNTs (red, with mainly (6,5)-chirality) at different excitation power (n=3, mean ± SD). The fit is a linear equation and the dotted fit (black) represents the extrapolation of the diffusion constant ($D_0$) to zero excitation power.*

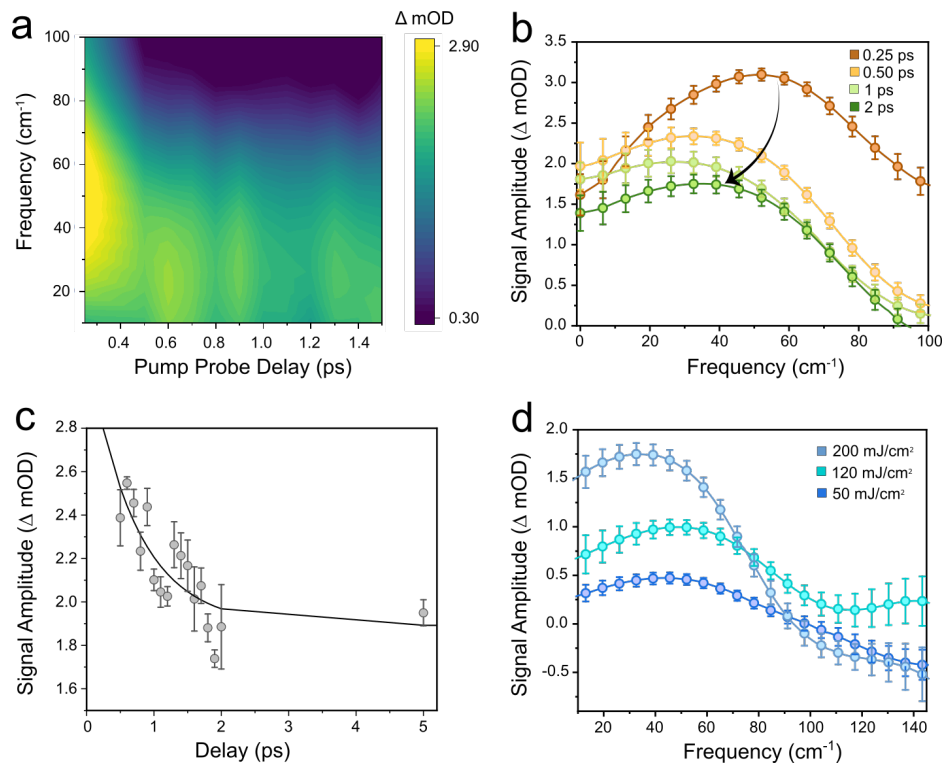

*Figure S9. Optical pump THz probe spectroscopy of DOC-SWCNTs in water: Temporal Dynamics and Fluence Dependence. a) Difference in THz absorption (absorption with light minus without light) of an aqueous DOC-SWCNTs solution as a function of time delay. b) THz spectra (corresponding to vertical slices of the map at representative pump-probe time delays (0.25, 0.5, 1, and 2 ps). The arrow serves as a guide to the eye. c) Plot of the maximum amplitudes of Δ mOD at a given time delay. The fitted exponential is also shown as a solid line, yielding a decay time of 0.71 ± 0.24 ps. d) Comparison of transient THz spectra at a pump-probe delay of 2 ps for distinct fluences (200, 120, and 50 mJ/cm$^2$).*

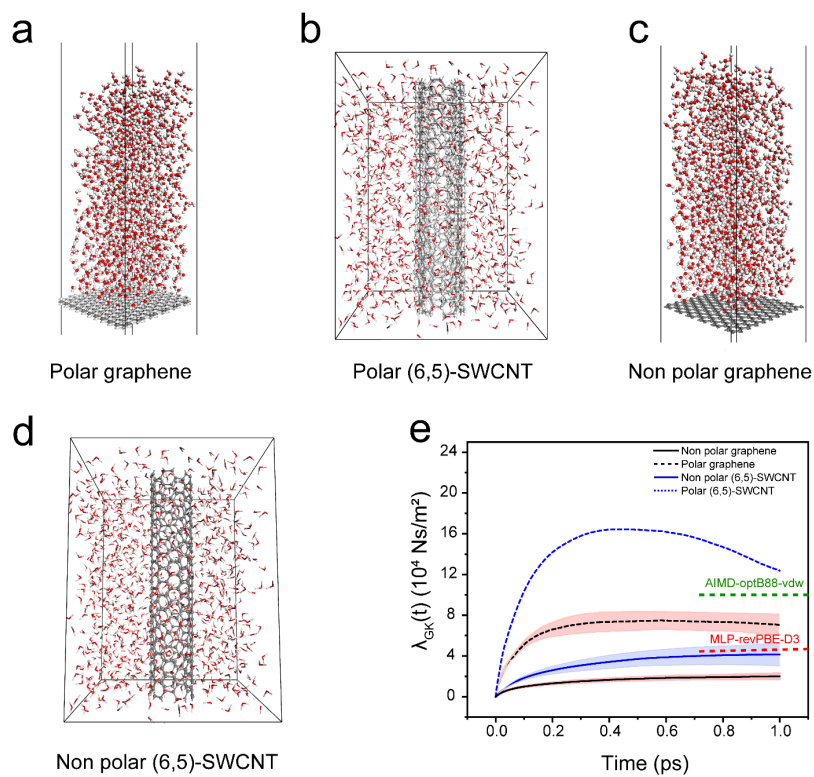

***Figure S10. Interfacial friction of graphene and (6,5)-SWCNTs with water.*** *a) Polar graphene-water interface modelled with IFF-CVFF. b) Polar SWCNT-water interface modeled with IFF-CVFF. In the polar models, a carbon atom (grey balls) is attached to two additional negatively virtual atoms (white balls attached to carbon atoms in panels a and b) to mimic a π electron cloud. c) Nonpolar graphene-water interface modeled with CVFF. d) Non-polar SWCNT-water interface modeled with CVFF. In the nonpolar models, a carbon atom (grey balls) is attached only to an adjacent carbon atom and thus it has no coulombic interaction with the solvent. In panels, a and c, the graphene water interface system is made non-periodic by leaving vacuum space inside the 3D periodic simulation box. In panel b and d, there is no water inside the nanotube and the system is 3D periodic. e) Green Kubo friction coefficient ($\lambda_{GK}$) for the polar and nonpolar graphene and SWCNTs.*

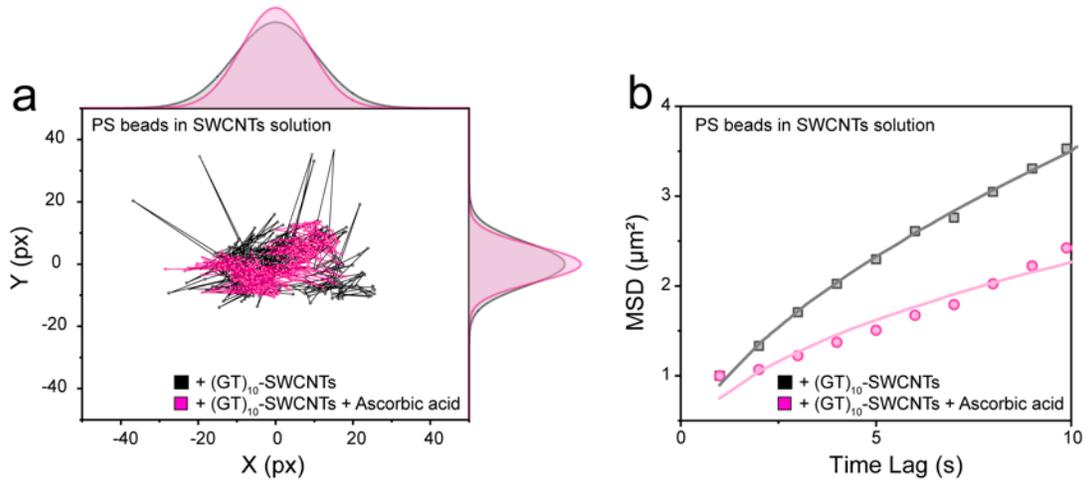

*Figure S11. **Macroscopic viscosity changes by quantum friction.** a) Example of single-particle trajectories of polystyrene (PS) beads (5 µm) diffusing in a PBS buffer solution containing $(GT)_{10}$-SWCNTs with and without ascorbic acid. Gaussian histogram plots illustrate the x,y positions. b) Ensemble time-averaged mean squared displacement (MSD) plots for PS beads in both conditions were examined (20 PS beads tracked over 399 frames), revealing slower diffusion in the presence of more excitons (ascorbic acid) (Table ST3).*

## FCS fit parameters

**Table ST1**. **$(GT)_{10}$-SWCNT FCS fitting parameters with increasing excitation power.** Off-state fraction *(T)*, off state lifetime $(\tau_t)$ and stretching exponent *(β)* as obtained from fitting of PL correlation curves recorded at various excitation powers (mean ± SD, n=3).

| Power (µW) | $\tau_d$ (ms) | D (µm²/s) | $\tau_t$ (ms) | T | β | N | Brightness |
|---|---|---|---|---|---|---|---|
| 10 | 14 ± 1.02 | 1.60 ± 0.07 | 2.0 ± 0.15 | 0.28 ± 0.008 | 0.80 ± 0.012 | 3 ± 0.31 | 1138 ± 400 |
| 20 | 16 ± 0.89 | 1.50 ± 0.04 | 2.0 | 0.28 | 0.72 ± 0.021 | 4 ± 0.61 | 1485± 310 |
| 40 | 18 ± 1.00 | 1.38 ± 0.07 | 2.0 | 0.28 | 0.61 ± 0.051 | 4.8 ± 0.49 | 2306 ± 256 |
| 60 | 20 ± 1.27 | 1.11 ± 0.04 | 2.0 | 0.28 | 0.52 ± 0.033 | 6.3 ± 0.25 | 3280 ± 166 |
| 80 | 22 ± 1.00 | 1.01 ± 0.09 | 2.0 | 0.28 | 0.48 ± 0.051 | 7.4 ± 0.32 | 4151 ± 298 |
| 90 | 24 ± 1.21 | 0.92 ± 0.06 | 2.0 | 0.28 | 0.47 ± 0.026 | 9.3 ± 0.21 | 4500 ± 300 |

**Table ST2**. **$(GT)_{10}$-SWCNTs in $D_2O$ (1nM) FCS fitting parameters with increasing excitation power.** (mean ± SD, n=3)

| Power (µW) | $\tau_d$ (ms) | D (µm²/s) | $\tau_t$ (ms) | T | β | Brightness | N |
|---|---|---|---|---|---|---|---|
| 10 | 30 ± 0.21 | 0.83 ± 0.008 | 1.0 ± 0.056 | 0.20 ± 0.006 | 0.60 ± 0.001 | 784 ± 11 | 1.5 ± 0.09 |
| 20 | 31 ± 0.49 | 0.81 ± 0.007 | 1.0 | 0.20 | 0.59 ± 0.002 | 869± 24 | 2.0 ± 0.23 |
| 40 | 30 ± 0.59 | 0.79 ± 0.01 | 1.0 | 0.20 | 0.58 ± 0.003 | 1033 ± 45 | 2.1 ± 0.12 |
| 60 | 31 ± 0.21 | 0.81 ± 0.01 | 1.0 | 0.20 | 0.58 ± 0.002 | 1195 ± 52 | 2.5 ± 0.14 |
| 80 | 31 ± 0.43 | 0.81 ± 0.005 | 1.0 | 0.20 | 0.57 ± 0.001 | 1378 ± 54 | 3.1 ± 0.12 |
| 90 | 31 ± 0.32 | 0.81 ± 0.006 | 1.0 | 0.20 | 0.57 ± 0.001 | 1474 ± 74 | 3.1 ± 0.24 |

*Table ST3.* **(GT)$_{10}$-SWCNTs in 80% glycerol (1nM).** FCS fitting parameters with increasing excitation power (mean ± SD, n=3).

| Power (µW) | $\tau_d$ (ms) | D (µm$^2$/s) | $\tau_t$ (ms) | T | β | Brightness | N |
|---|---|---|---|---|---|---|---|
| 10 | 43 ± 1.4 | 0.66 ± 0.001 | 1 ± 1.2 | 0.10 ± 0.009 | 0.86 ± 0.008 | 400 ± 33 | 2.0 ± 0.09 |
| 90 | 43 ± 1.2 | 0.66 ± 0.002 | 1 | 0.10 | 0.82 ± 0.01 | 582 ± 31 | 3.0 ± 0.12 |

*Table ST4.* **(GT)$_{10}$-SWCNTs in 20% glycerol (1nM).** FCS fitting parameters with increasing excitation power (mean ± SD, n=3).

| Power (µW) | $\tau_d$ (ms) | D (µm$^2$/s) | $\tau_t$ (ms) | T | β | Brightness | N |
|---|---|---|---|---|---|---|---|
| 10 | 15 ± 1.20 | 1.67 ± 0.02 | 2.2 ± 0.12 | 0.30 ± 0.005 | 0.70 ± 0.010 | 1000 ± 200 | 4 ± 0.40 |
| 40 | 20 ± 1.00 | 1.25 ± 0.03 | 2.2 | 0.30 | 0.60 ± 0.012 | 1258 ± 160 | 6 ± 0.32 |
| 80 | 23 ± 1.10 | 1.09 ± 0.03 | 2.2 | 0.30 | 0.58 ± 0.024 | 3125 ± 193 | 6.3 ± 0.27 |
| 90 | 25 ± 1.00 | 1.00 ± 0.04 | 2.2 | 0.30 | 0.52 ± 0.029 | 4210 ± 138 | 7.1 ± 0.31 |

**Table ST5.** Concentration dependence of FCS data of 1nM **(GT)$_{10}$-SWCNTs exposed to riboflavin** at 20 µW laser power.

| Riboflavin (µM) | $\tau_d$ (ms) | D (µm$^2$/s) | $\tau_t$ (ms) | T | β | N | Brightness |
|---|---|---|---|---|---|---|---|
| 0 | 16 ± 1.01 | 1.45 ± 0.04 | 2.0 ± 0.18 | 0.28 ± 0.008 | 0.72 ± 0.01 | 4 ± 0.98 | 1098 ± 70 |
| 0.1 | 15 ± 0.80 | 1.63 ± 0.08 | 2.0 | 0.28 | 0.70 ± 0.02 | 4.8 ± 0.78 | 1028 ± 52 |
| 0.5 | 12 ± 1.07 | 1.89 ± 0.09 | 2.0 | 0.28 | 0.68 ± 0.01 | 5.8 ± 0.32 | 872 ± 45 |
| 1 | 10 ± 0.50 | 2.37 ± 0.08 | 2.0 | 0.28 | 0.55 ± 0.01 | 6.5 ± 0.45 | 753 ± 38 |
| 2 | 8 ± 0.41 | 2.98 ± 0.14 | 2.0 | 0.28 | 0.45 ± 0.01 | 8.0 ± 0.38 | 510 ± 70 |

**Table ST6.** Concentration dependence of FCS data of 1nM **(GT)$_{10}$-SWCNTs exposed to ascorbic acid** at 20 µW laser power.

| Ascorbic (µM) | $\tau_d$ (ms) | D (µm$^2$/s) | $\tau_t$ (ms) | T | β | N | Brightness |
|---|---|---|---|---|---|---|---|
| 0 | 16 ± 1.00 | 1.45 ± 0.04 | 2.0 ± 0.18 | 0.28 ± 0.008 | 0.72 ± 0.08 | 4 ± 0.12 | 1084 ± 213 |
| 5 | 18 ± 1.20 | 1.39 ± 0.05 | 2.0 | 0.28 | 0.73 ± 0.09 | 4.1 ± 82 | 1443 ± 243 |
| 20 | 20 ± 1.57 | 1.24 ± 0.03 | 2.0 | 0.28 | 0.78 ± 0.04 | 3.5 ± 0.42 | 2410 ± 312 |
| 50 | 24 ± 1.20 | 1.04 ± 0.06 | 2.0 | 0.28 | 0.82 ± 0.06 | 3.0 ± 21 | 3726 ± 501 |
| 100 | 37 ± 1.40 | 0.67 ± 0.04 | 2.0 | 0.28 | 0.88 ± 0.03 | 2.0 ± 0.82 | 6000 ± 458 |

**Table ST7. DOC-SWCNTs (1 nM) FCS fitting parameters with increasing excitation power**. (mean ± SD, n=3).

| Power (µW) | $\tau_d$ (ms) | D (µm$^2$/s) | $\tau_t$ (ms) | T | β | Brightness | N |
|---|---|---|---|---|---|---|---|
| 10 | 26 ± 1.0 | 0.96 ± 0.03 | 9.0 ± 0.21 | 0.50 ± 0.002 | 0.74 ± 0.01 | 1272 ± 100 | 1.8 ± 0.21 |
| 20 | 27 ± 2.1 | 0.91 ± 0.01 | 9.0 | 0.50 | 0.71 ± 0.02 | 2250 ± 190 | 2.7 ± 0.13 |
| 40 | 32 ± 1.3 | 0.78 ± 0.02 | 9.0 | 0.50 | 0.68 ± 0.01 | 2987 ± 195 | 2.1 ± 0.10 |
| 60 | 36 ± 2.5 | 0.69 ± 0.01 | 9.0 | 0.50 | 0.62 ± 0.02 | 4265 ± 100 | 3.2 ± 0.21 |
| 80 | 41 ± 2.8 | 0.60 ± 0.03 | 9.0 | 0.50 | 0.52 ± 0.01 | 4885 ± 134 | 4.2 ± 0.12 |
| 90 | 44 ± 2 | 0.57 ± 0.02 | 9.0 | 0.50 | 0.51 ± 0.01 | 5721 ± 145 | 5.0 ± 0.18 |

**Table ST8. DOC-SWCNTs with Nitro-Aryl sp$^3$ defect (1nM) FCS fitting parameters with increasing excitation power**. (mean ± SD, n=3).

| Power (µW) | $\tau_d$ (ms) | D (µm$^2$/s) | $\tau_t$ (ms) | T | β | Brightness | N |
|---|---|---|---|---|---|---|---|
| 10 | 19 ± 1.1 | 1.32 ± 0.01 | 4.0 ± 0.12 | 0.30 ± 0.01 | 0.80 ± 0.01 | 300 ± 10 | 2.3 ± 0.42 |
| 20 | 19 ± 1.1 | 1.31 ± 0.02 | 4.0 | 0.30 | 0.80 ± 0.01 | 312 ± 10 | 2.5 ± 0.21 |
| 40 | 19.5 ± 1.0 | 1.28 ± 0.01 | 4.0 | 0.30 | 0.78 ± 0.02 | 331 ± 9 | 3.0 ± 0.12 |
| 60 | 20 ± 1.2 | 1.25 ± 0.02 | 4.0 | 0.30 | 0.78 ± 0.01 | 368 ± 23 | 3.1 ± 0.21 |
| 80 | 20 ± 1.0 | 1.22 ± 0.01 | 4.0 | 0.30 | 0.78 ± 0.02 | 412 ± 17 | 4.0 ± 0.24 |
| 90 | 20.5 ± 1.1 | 1.22 ± 0.01 | 4.0 | 0.30 | 0.78 ± 0.01 | 425 ± 19 | 4.5 ± 0.14 |

**Table ST9. SC-SWCNTs (1nM) FCS fitting parameters with increasing excitation power**. (mean ± SD, n=3).

| Power (µW) | $\tau_d$ (ms) | D (µm$^2$/s) | $\tau_t$ (ms) | T | β | Brightness | N |
|---|---|---|---|---|---|---|---|
| 10 | 24 ± 1.20 | 1.00 ± 0.02 | 2.0 ± 0.04 | 0.60 ± 0.007 | 0.68 ± 0.02 | 1228 ± 100 | 2.5 ± 0.18 |
| 20 | 26 ± 1.40 | 0.96 ± 0.01 | 2.0 | 0.60 | 0.64 ± 0.02 | 1460 ± 98 | 3.1 ± 0.19 |
| 40 | 28 ± 2.30 | 0.89 ± 0.03 | 2.0 | 0.60 | 0.60 ± 0.01 | 1880 ± 143 | 4.0 ± 0.16 |
| 60 | 30 ± 1.40 | 0.83 ± 0.01 | 2.0 | 0.60 | 0.60 ± 0.01 | 2071 ± 120 | 5.0 ± 0.27 |
| 80 | 32 ± 1.60 | 0.78 ± 0.02 | 2.0 | 0.60 | 0.58 ± 0.02 | 2463 ± 131 | 5.6 ± 0.22 |
| 90 | 34 ± 1.20 | 0.73 ± 0.01 | 2.0 | 0.60 | 0.54 ± 0.01 | 2635 ± 140 | 6.0 ± 0.32 |

**Table ST10. SDBS-SWCNTs with sp$^3$ defect (1nM) FCS fitting parameters with increasing excitation power**. (mean ± SD, n=3).

| Power (µW) | $\tau_d$ (ms) | D (µm$^2$/s) | $\tau_t$ (ms) | T | β | Brightness | N |
|---|---|---|---|---|---|---|---|
| 10 | 18 ± 0.30 | 1.30 ± 0.02 | 2.3 ± 0.08 | 0.60 ± 0.002 | 0.86 ± 0.01 | 546 ± 23 | 3.0 ± 0.12 |
| 20 | 20 ± 0.50 | 1.24 ± 0.01 | 2.3 | 0.60 | 0.84 ± 0.01 | 581 ± 12 | 3.8 ± 0.19 |
| 40 | 21 ± 0.21 | 1.16 ± 0.01 | 2.3 | 0.60 | 0.80 ± 0.02 | 683 ± 33 | 4.0 ± 0.10 |
| 60 | 23 ± 0.42 | 1.06 ± 0.01 | 2.3 | 0.60 | 0.78 ± 0.01 | 803 ± 24 | 5.7 ± 0.09 |
| 80 | 25 ± 0.45 | 1.02 ± 0.02 | 2.3 | 0.60 | 0.78 ± 0.01 | 902 ± 32 | 5.8 ± 0.10 |
| 90 | 26 ± 0.34 | 1.00 ± 0.01 | 2.3 | 0.60 | 0.76 ± 0.01 | 956 ± 21 | 6.0 ± 0.10 |

**Table ST11. DOC-(6,4)-SWCNTs (1nM) with power**. FCS fitting parameters with increasing excitation power (mean ± SD, n=3).

| Power (µW) | $\tau_d$ (ms) | D (µm$^2$/s) | $\tau_t$ (ms) | T | β | Brightness | N |
|---|---|---|---|---|---|---|---|
| 10 | 27 ± 0.7 | 0.93 ± 0.01 | 2.5 ± 0.12 | 0.70 ± 0.02 | 0.77 ± 0.01 | 3533 ± 121 | 1.0 ± 0.10 |
| 20 | 28 ± 0.8 | 0.88 ± 0.02 | 2.5 | 0.70 | 0.77 ± 0.008 | 4221 ± 187 | 2.0 ± 0.13 |
| 60 | 35 ± 1.0 | 0.71 ± 0.01 | 2.5 | 0.70 | 0.72 ± 0.01 | 6550 ± 196 | 2.3 ± 0.12 |
| 90 | 41 ± 0.9 | 0.61 ± 0.01 | 2.5 | 0.70 | 0.69 ± 0.01 | 9122 ± 132 | 3.5 ± 0.21 |

**Table ST12. Linear fit quality for the various functionized SWCNTs for the power dependent diffusion constant.**

| Surface modification | R$^2$ |
|---|---|
| (GT)$_{10}$-SWCNTs | 0.996 |
| SDBS-SWCNTs | 0.967 |
| SC-SWCNTs | 0.997 |
| DOC-SWCNTs | 0.990 |
| DOC-(6,4)-SWCNTs | 0.995 |

**Table ST13. MSD fitting parameters**. Diffusion constant D and power exponent n were obtained from a linear regression in log space.

| Sample | α | D (µm$^2$/s) | R$^2$ |
|---|---|---|---|
| Rh B PS beads with (GT)$_{10}$ –SWCNT | 0.59 | 0.89 | 0.99 |
| Rh B PS beads with (GT)$_{10}$ –SWCNT and ascorbic acid | 0.47 | 0.74 | 0.95 |

**Table ST14. Polar and Non-polar model parameters.** Table of parameters for classical MD of a polarizable and nonpolar CNT and flexible SCPE water model along with equation for total energy (Etotal) comprising kinetic (KE) and potential energy (PE) terms. All the models are also included as part of the supporting files.

| $E_{tot}=E_{KE}+E_{PE}$ where $E_{PE}=E_{Bond} + E_{Angle} + E_{dihedral} + E_{Improper} + E_{vdw} + E_{Coulomb}$ | | | |
|---|---|---|---|
| $E_{Bond}=\mathbf{k}(r_{ij}-\mathbf{r_0})^2$; $E_{Angle}= \mathbf{k}(\Theta_{ijk}-\mathbf{\Theta_0})^2$; $E_{dihedral}=\mathbf{k}[1+\mathbf{d}\cos(\mathbf{n}\Phi_{ijkl})]$ ; $E_{Improper}= \mathbf{k}[1+\mathbf{d}\cos(\mathbf{n}\Phi_{ijkl})]$ | | | |
| $E_{Coulomb}= \mathbf{k}\mathbf{q_i}\mathbf{q_j}/r_{ij}^2$; $E_{vdw}= \mathbf{\varepsilon}[(\mathbf{\sigma_0}/r_{ij})^{12}-2(\mathbf{\sigma_0}/r_{ij})^6]$ | | | |
| **IFF-CVFF Polarizable Graphene/CNT Parameters** | | | |
| Mass (amu)[1] | C 10.011150 | D 1.000000 | |
| Bond<br>k: kcal/mol, $r_0$: Å | C-C<br>(k=480, $r_0$=1.558) | C-D<br>(k=150, $r_0$=0.60) | D-D[2]<br>(k=0, $r_0$=1.558) |
| Angle<br>k: kcal/mol, $\Theta_0$: deg. | C-C-C<br>(k=60, $\Theta_0$=120) | C-C-D<br>(k=25, $\Theta_0$=90) | D-C-D<br>(k=25, $\Theta_0$=180) |
| Dihedral<br>k: kcal/mol, r0: Å | C-C-C-C | (k=24, d= -1, n= 2) | |
| Non-Bond[3]<br>q:e, $\sigma_0$:Å, ε:kcal/mol | C<br>(q = 0.8, σ = 4.167, ε = 0.025) | | D<br>(q = -0.4, σ = 1.800, ε = 0.045) |
| **CVFF Non-polar Graphene/CNT Parameters** | | | |
| Mass (amu)[1] | C 12.011150 | | |
| Bond (k: kcal/mol, $r_0$: Å) | C-C (k=480, $r_0$=1.340) | | |
| Angle (k: kcal/mol, $\Theta_0$: deg.) | C-C-C (k=90, $\Theta_0$=120) | | |
| Dihedral (k: kcal/mol ) | C-C-C-C (k=3, d= -1, n= 2) | | |
| Improper (k: kcal/mol) | C-C-C-C (k=0.37, d= -1, n= 2) | | |
| Non-Bond[3] (q:e, $\sigma_0$:Å, ε:kcal/mol) | C (q = 0.0, σ = 4.060, ε = 0.148) | | |
| **Flexible SPC Water Parameters** | | | |
| Mass (amu)[1] | O 15.999400 | | H 1.007970 |
| Bond (k: kcal/mol, $r_0$: Å) | O-H (k=540.634, $r_0$=0.960) | | |
| Angle (k: kcal/mol, $\Theta_0$: deg.) | H-O-H (k=50, $\Theta_0$=104.5) | | |
| Non-Bond[3]<br>(q: e, $\sigma_0$: Å, ε: kcal/mol) | O<br>(q=-0.82, σ=3.553, ε=0.155) | | H<br>(q=+0.41, σ=0.242, ε=0.025) |

[1]C=Carbon, D=Dummy atom, O=Oxygen, and H=Hydrogen, [2]Bonded interactions are added to the dummy atoms in the polarizable model to avoid calculation of 1-4 interactions to increase the flexibility of the dummy atoms in the polarizable graphite/CNT model even though some of the bonds and angles constants are 0. [3]Geometric mixing rule is applied to evaluate the ij Lennard Jonesd interaction parameters from pure component ii and jj interactions.